\documentclass{ws-procs961x669}            

\usepackage[utf8]{inputenc}
\usepackage{amsmath}
\usepackage{color}
\usepackage{polski}
\usepackage[breaklinks]{hyperref}
\usepackage{natbib}

\begin{document}
\title{Finite Action Principle and wormholes.}

\author{Jan Chojnacki${}^{1}$}

\address{${}^{1}$Faculty of Physics, University of Warsaw, ul. Pasteura 5, 02-093 Warsaw, Poland}

\author{ Jan Kwapisz${}^{1,2}$}

\address{${}^{1}$Faculty of Physics, University of Warsaw, ul. Pasteura 5, 02-093 Warsaw, Poland \\
${}^{2}$CP3-Origins, University of Southern Denmark, Campusvej 55, DK-5230 Odense M, Denmark}

\begin{abstract}
In this work, we elaborate on the finite action for wormholes in higher derivative theories. Both non-traversable and traversable wormholes in theories with higher curvature invariants posses finite action. 

\end{abstract}

\keywords{Horava gravity; Wormholes; finite action}

\bodymatter
\section{Introduction}
The Finite Action Principle (FAP) proposing that the physical spacetimes are those for which the action is finite. This comes reasoning comes from the fact that in Euclidian path integral the weighting factor is $e^{-S[\Phi]}$. Hence lesser the action the more it contributes to the path integral. On the other hand in the Lorentztian signature the more varying actions contribute less. Hence for the actions neighbouring the infinite one the variation is significant and the neighbouring ones do cancel each other, see \cite{borissova2020blackhole}. \\
This notion has recently received significant attention in the literature, see \cite{finiteaction,Jonas:2021xkx,Chojnacki:2021ves} for the cosmological considerations and \cite{borissova2020blackhole,Chojnacki:2021ves} for the black hole ones. Since it is expected that the quantum gravity should resolve the black-hole singularity problem, one may ask which of the microscopic actions remain finite for non-singular black holes and conversely interfere destructively for the singular ones. This we shall call the finite action selection principle. Only after the inclusion of higher-curvature operators, beyond the Einstein-Hilbert term, such selection principle can be satisfied \cite{borissova2020blackhole}. Furthermore in asymptotic safety, the quantum corrections to the Newtonian potential eliminate the classical-singularity \cite{Bosma:2019aiu}. \\
These findings suggest that by taking into account the higher curvatures one can resolve the singularities in black holes. Yet, an issue with the higher-curvature theory of quantum gravity is the existence of the particles with the negative mass-squared spectrum, known as \textit{ghosts}, which makes the theory non-unitary.
In this article, we explore possible resolution, namely, we investigate Horava-Lifshitz (H-L) gravity \cite{Ho_ava_2009}, where the Lorentz Invariance (LI) is broken at the fundamental level (see \cite{Wang_2017} for a comprehensive progress report on this subject). Kinetic terms are first order in the time derivatives, while higher spatial curvature scalars regulate the UV behavior of the gravity. In our investigation we follow Horava \cite{Ho_ava_2009} and assume that Wick rotation is well defined. This can be motivated by the lack of the higher time-like curvature invariants. In the usual higher derivative constructions the existence of massive poles in the propagator makes the Wick rotation troublesome \cite{Donoghue:2019fcb}. Conversely to the latter, the new poles are massless and non-tachyonic for suitable choice of parameters \cite{Bemfica:2011xv}. Furthermore, the existence of the foliation supports that claim. For the discussion of Minkowski path integral and FAP, see \cite{borissova2020blackhole}. \\
In the previous article, we discussed the possibility that the Finite Action arguments applied to the projectable H-L gravity result in a flat, homogeneous, UV-complete, and ghost-free beginning of the universe \cite{Chojnacki:2021ves}. However there is no known regular black hole solutions in HL gravity.  Here we discuss the wormholes solutions in context of the Finite Action Principle in HL gravity as well as in the $f(R)$ theories. Interestingly the stable, traversable wormholes solutions are known only in the higher derivative gravities \cite{Duplessis_2015} (without exotic matter), so there seems to be a wormhole/non-singular BH trade-off after taking into account the Finite Action Principle. \\
\section{Horava-Lifshitz gravity}
In the Horava-Lifshitz gravity, space and time are scaled in a non-equivalent way. Diffeomorphism invariance is broken by the foliation of the 4-dimensional spacetime into 3-dimensional hypersurfaces of constant time, called leaves, making the theory power-counting renormalizable (see also the renormalization group studies of the subject \cite{DOdorico:2014tyh,DOdorico:2015pil,Barvinsky:2017kob}). The remaining symmetry respects transformations:
\begin{align}
    t\xrightarrow{}\xi_0(t),\quad x^i\xrightarrow{}\xi^i(t,x^k),
\end{align}
and is often referred to as the foliation-preserving diffeomorphism, denoted by Diff$(M,\mathcal{F})$. The diffeomorphism invariance is still present on the leaves. The four-dimensional metric may be expressed in the Arnowitt-Deser-Misner (ADM) \cite{Arnowitt_2008} variables:
\begin{align}
\label{ADM}
    (N,N^i,\,^{(3)}g_{ij}),
\end{align}
where $N,\, N^i,\,\,^{(3)}g_{ij}$ denote respectively the lapse function, shift vector, and 3-dimensional induced metric on the leaves. The theory is constructed from the following quantities:
\begin{align}
    \,^{(3)}R_{ij},\quad K_{ij},\quad a_i,\quad \,^{(3)}\nabla_i,
\end{align}
where $\,^{(3)}R_{ij}$ is the 3-dimensional Ricci curvature tensor, $\,^{(3)}\nabla_i$ is the covariant derivative constructed from the 3-dimensional metric $\,^{(3)}g_{ij}$, and $a_i:=\frac{N,_i}{N}$. Extrinsic curvature $K_{ij}$ is the only object, invariant under general spatial diffeomorphisms containing exactly one time derivative of the metric tensor $\,^{(3)}g_{ij}$:
\begin{align}
    K_{ij}=\frac{1}{2 N} \left(\frac{\partial \,^{(3)}g_{ij}}{\partial t}- \,^{(3)}\nabla_i N_j -\,^{(3)}\nabla_j N_i   \right).
\end{align}
Quantities (\ref{ADM}) are tensor/vectors with respect to Diff($M,\mathcal{F}$) possessing the following mass dimensions:
\begin{align}
    [\,^{(3)}R_{ij}]=2,\quad [K_{ij}]=3,\quad [a_i]=1,\quad [\,^{(3)}\nabla_i]=1.
\end{align}
One may use (\ref{ADM}) to construct, order by order, scalar terms appearing in the Lagrangian of the theory.  Following \cite{Wang_2017,Maier_2017} the action of the Horava gravity takes the form: 
\begin{align}
\label{ProjectableAction}
S_g = \zeta^2 \int dt dx^{3} N\sqrt{^{(3)}g} \left(\mathcal{K}-V\right),
\end{align}
where $\mathcal{K} = K_{ij}K^{ij}-\lambda K^2$ with $K=K_{ij}\,^{(3)}g^{ij} $, $\,^{(3)}g$ denotes the determinant of the 3-dimensional metric and $\zeta^2 = 1/16 \pi G$.
It may be expressed as the difference of the kinetic and potential part
$\mathcal{L}=\mathcal{K}-V$ with
$\mathcal{K}= \left(K_{ij}K^{ij}-\lambda K^2 \right)$. At the 6th order, the potential part of the lagrangian contains over 100 terms \cite{Wang_2017}. The immense number of invariants is limited by imposing further symmetries. One possible restriction for the potential comes from the projectability condition $N=N(t)$, then terms proportional to $a_i\equiv 0$ vanish. Up to the sixth order (compatible with power counting renormalizability), the potential $V$ restricted by the projectability condition is given by:
\begin{align}
\label{eq:potentialproj}
    V=2\Lambda \zeta^2-\,^{(3)}R+\frac{1}{\zeta^2} \left(g_2 \,^{(3)}R^2+g_3\,^{(3)}R^{ij}\,^{(3)}R_{ij}\right) \nonumber \\
    +\frac{1}{\zeta^4} \left(g_4{}^{(3)}R^3+g_5\,^{(3)}R\,^{(3)}R^{ij}\,^{(3)}R_{ij}+g_6\,^{(3)}R^{i}_j\,^{(3)}R^{j}_k\,^{(3)}R^{k}_i\right),\nonumber \\
     +\frac{1}{\zeta^4} \left(g_7{}^{(3)}R\nabla^2{}^{(3)}R + g_8 (\nabla_i {}^{(3)}R_{jk})(\nabla^i {}^{(3)}R^{jk}) \right),
\end{align}
where $\Lambda$ is the cosmological constant and $\alpha_{ij}$ are the coupling constants. For our purposes, we drop terms containing covariant derivatives $\,^{(3)}\nabla_i$. One should also mention that this \emph{minimal theory} \cite{Horava:2011gd} suffers from the existence of spin 0 graviton, which is unstable in the IR. Various solutions to this problem have been proposed. One can add the additional local $U(1)$ symmetry \cite{Wang_2017,Horava:2010zj}. Then by the introduction of new fields prevents the zero-mode from propagating. On the other hand, one can drop the projectability condition $a_i=0$ and include the terms containing $a_i$ in the potential term:
\begin{align}
    V = 2\Lambda \zeta^2 -\,^{(3)}R - \beta_0 a_i a^i + \sum_{n=3}^6 \mathcal{L}_V^{(n)},
\end{align}
then for the spin-0 mode to be stable one requires $0<\beta_0<2$ \cite{Blas:2010hb,Carloni:2010ji}. 
\section{Black holes and wormholes}
In this section, we show that H-L gravity satisfies the Finite Action selection principle for the microscopic action of quantum gravity \cite{borissova2020blackhole}. We study both the solutions of H-L gravity and the known, off-shell BH spacetimes, due to the fact that there are no known regular BH solutions in the HL gravity. Keep in mind that a metric does not need to be a solution to the equations of motion to enter the path integral. We require singular black-hole metrics to interfere destructively, while the regular ones with finite action contribute to the probability amplitudes. We broaden this analysis by studying the wormhole solutions. 
\section{Wormholes and finite action principle}
Here, we take the first step in the direction of the investigations of the consequences of the Finite Action Principle in the context of wormholes (WH). The wormholes may be characterized in two classes: traversable and non-traversable. The traversable WH, colloquially speaking, are such that one can go through it to the other side, see \cite{Morris:1988cz} for specific conditions.
The pioneering Einstein-Rosen bridge has been found originally as a non-static, non-traversable solution to GR. The traversable solutions are unstable, however, they might be stabilized by an exotic matter or inclusion of the higher curvature scalar gravity \cite{Duplessis_2015}. This is important in the context of finite action since usually the divergences of black holes do appear in the curvature squared terms. Hence, due to the inclusion of the higher-order terms in the actions, the traversable wormholes are solutions to the equations of motions without the exotic matter. The exemplary wormhole spacetimes investigated here are the Einstein-Rosen bridge proposed in \cite{PhysRev.48.73}, the Morris-Thorne (MT) wormhole \cite{Morris:1988cz}, the traversable exponential metric wormhole \cite{Boonserm_2018} and the wormhole solution discussed in the H-L gravity \cite{Botta_Cantcheff_2010}. All of them have a finite action. Here, we shall discuss the exponential metric WH. The conclusions for the other possible wormholes are similar and we discuss them in the Appendix. For the exponential metric WH, the line element is given by:
\begin{align}
    ds^2=-e^{-\frac{2M}{r}}dt^2+e^{\frac{2M}{r}}\left(dr^2+r^2 d\Omega^2 \right).
\end{align}
 This spacetime consists two regions: ``our universe'' with $r>M$ and the ``other universe" with $r<M$. $r=M$ corresponds to the wormhole's throat. The spacial volume of the ``other universe" is infinite when $r\xrightarrow{}0$. Such volume divergence is irrelevant to our discussion since it describes large distances in the "other universe". Hence, we further consider only $r\geq M$. 
The resulting Ricci and Kretschmann scalars calculated in \cite{Boonserm_2018} and the measure are non-singular everywhere:
\begin{align}
    R&=-\frac{2M^2}{r^4} e^{\frac{-2M}{r}},\nonumber\\
    R_{\mu\nu\sigma\rho}R^{\mu\nu\sigma\rho}&=\frac{4M^2 (12r^2-16Mr+7M^2)}{r^8}e^{-4\frac{M}{r}},
\end{align}
resulting in the finite action for the Stelle gravity. Similarly for the H-L gravity:
\begin{align}
    {}^{(3)}R=R,\quad K^2=K_{ij}K^{ij}=0,\nonumber\\
    {}^{(3)}R_{ij} {}^{(3)}R^{ij}=\frac{2 M^2 (M^2-2 M r+3 r^2)}{r^8}e^{-\frac{4 M}{r}}.
\end{align}
\paragraph{Einstein-Rosen bridge}
The Einstein-Rosen (E-R) bridge smoothly glues together two copies of the Schwarzschild spacetime: black-hole and the white-hole solutions corresponding to the 
positive and negative coordinate $u$.
Metric tensor of the Einstein-Rosen wormhole proposed in \cite{PhysRev.48.73} and discussed in e.g. \cite{Katanaev_2014} is given by:
\begin{align}
    ds^2=\frac{-u^2}{u^2+4M}dt^2+(u^2+4M)du^2+\frac{1}{4}(u^2+4M)d\Omega^2.
\end{align}
The E-R bridge is non-traversable and geodesically incomplete in $u=0$. This fact, however, does not impact the regularity of the curvature scalars.
The 4-dimensional Ricci scalar is:
\begin{align}
 R=\frac{2 \left(64 M^2+32 M u^2+4 u^4+u^2\right)}{\left(4 M+u^2\right)^3}.   
\end{align}
The second order curvature scalar $R_{\mu\nu}R^{\mu\nu}$ is:
\begin{align}
  \frac{4 \left(48 M^2+8 \left(4 M+u^2\right)^4+(32 M-1) \left(4 M+u^2\right)^2\right)}{\left(4 M+u^2\right)^6}.
\end{align}
Both of which integrated with the measure are non-singular:
\begin{align}
    \sqrt{g}=\frac{1}{4}u(4M+u^2).
\end{align}
The wormhole solutions analyzed in this paper generally yield the finite action in both GR and H-L. The finite Action Principle suggests, that in the quantum UV regime, singular black-hole spacetimes may be replaced with the regular wormhole solutions.
\paragraph{Morris-Thorne wormhole}
The MT wormhole is defined in the spherically symmetric, Lorentzian spacetime by the line element:
\begin{align}
    ds^2=-e^{2\Phi(r)}dt^2+\frac{dr^2}{1-\frac{b(r)}{r}}+r^2 d\Omega^2
\end{align}
where $\Phi(r)$ is known as the redshift and there are no horizons if it is finite. Function $b(r)$ determines the wormhole's shape. We choose $\Phi(r),\,b(r)$ to be:
\begin{align}
    \Phi(r)&=0,\quad
    b(r)=2M\left(1-e^{r_0-r}\right)+r_0 e^{r_0-r},
\end{align}
where $r_0$ is the radius of the throat of the wormhole, such that $b(r_0)=r_0$. 4-dimensional curvature scalars for this spacetime have been calculated in \cite{Mattingly_2020}. The Ricci curvature scalar is singular at $r=0$, however, the radial coordinate $r$ varies between $r_0>0$ and infinity:
\begin{align}
\label{RicciM-T}
    R=-2\left(2M-r_0\right)\frac{e^{r_0-r}}{r^2}.
\end{align}
The resulting $S_s=\int_{r_{UV}=r_0}^{r_{IR}} \sqrt{g}R$ function is divergent as $r_{UV}\xrightarrow{}r_0$ and cannot be expressed in terms of simple functions:
\begin{align}
    2(2M-r_0)\int_{r_{UV}=r_0}^{r_{IR}}\sqrt{\frac{r}{r-2M(1-e^{r_0-r})+r_0 e^{r_0-r}}}e^{r_0-r} dr.
\end{align}
However, this is only a coordinate singularity and one may get rid of it with a proper transformation.\\
Higher-order curvature scalars for Morris-Thorne wormhole are:
\begin{align}
    {}^{(3)}R&=\frac{2 b'(r)}{r^2}\nonumber\\
    {}^{(3)}R_{ij}{}^{(3)}R^{ij}&=\frac{3 r^2 b'(r)^2-2 r b(r) b'(r)+3 b(r)^2}{2 r^6},\nonumber\\
    {}^{(3)}R^i_{j}{}^{(3)}R^j_k {}^{(3)}R^k_i&=\frac{-9 r^2 b(r) b'(r)^2+5 r^3 b'(r)^3+15 r b(r)^2 b'(r)-3 b(r)^3}{4 r^9},
\end{align}
and integrated give action that is finite.
\paragraph{H-L wormhole}
Static spherically traversable symmetric wormholes have been constructed in \cite{Botta_Cantcheff_2010} in the H-L theory through the modification of the Rosen-Einstein spacetime:
\begin{align}
    ds^2=-N^2(\rho)dt^2+\frac{1}{f(\rho)}d\rho^2+(r_0+\rho^2)^2 d\Omega^2,
\end{align}
with additional $\mathbf{Z}_2$ symmetry with respect to the wormhole's throat. There are solutions with $\lambda=1$ asymptotically corresponding to the Minkowski vacuum. Explicitly we have:
\begin{align}
    f=N^2&=1+\omega (r_0+\rho^2)^2\nonumber\\&-\sqrt{(r_0+\rho^2)\left(\omega^2(r_0+\rho^2)^3+4\omega M\right)}.
\end{align}
Radius of the wormhole's throat is given by $r_0$. The parameters $\omega,$ and $M$ are connected to the coupling constants in H-L action. See \cite{Botta_Cantcheff_2010} for their explicit form.
Ricci scalar of the H-L wormhole invariants are given by
\begin{align}
    {}^{(3)}R&=-\frac{1}{{(\rho
   ^2+r_0){}^2}}\Big[2 (-10 \rho ^2 \sqrt{\omega  (\rho ^2+r_0) (4 M+\omega  (r_0+\omega ^2){}^3)}\nonumber\\&-4 r_0 \sqrt{\omega  (\rho ^2+r_0) (4
   M+\omega  (r_0+\omega ^2){}^3)}+16 \rho ^6 \omega +8 \rho ^2\nonumber\\&+36 \rho ^4 r_0 \omega +24 \rho ^2 r_0^2 \omega +4 r_0^3 \omega +4 r_0-1)\Big],\nonumber\\
   {}^{(3)}R_{ij}{}^{(3)}R^{ij}&= \frac{1}{(\rho ^2+r_0)^4}\left\{2 (-7 \rho ^2 \sqrt{\omega  (\rho ^2+r_0) (4 M+\omega  (r_0+\omega ^2){}^3)}-2 r_0 \sqrt{\omega  (\rho ^2+r_0) (4 M+\omega
    (r_0+\omega ^2){}^3)}\right. \nonumber\\& +10 \rho ^6 \omega 
    +6 \rho ^2+22 \rho ^4 r_0 \omega+14 \rho ^2 r_0^2 \omega +2 r_0^3 \omega +2 r_0-1){}^2+4 M \nonumber\\&
    +\frac{4 (\rho ^2+r_0)}{\omega  (4 M+\omega  (r_0+\omega ^2){}^3)}\left[\rho ^2 \omega  (-4 (\rho ^2+r_0) \sqrt{\omega  (\rho ^2+r_0) (4 M+\omega  (r_0+\omega ^2){}^3)}+\omega 
   (r_0+\omega ^2){}^3)\right]\nonumber\\&
   \left.-2 \sqrt{\omega  (\rho ^2+r_0) (4 M+\omega  (r_0+\omega ^2){}^3)} \left[-\sqrt{\omega  (\rho
   ^2+r_0) (4 M+\omega  (r_0+\omega ^2){}^3)}+\omega  (\rho ^2+r_0){}^2+1)\right]{}^2\right\}
\end{align}
The kinetic terms with $K_{ij}=0$ are vanishing, while the spacial Ricci scalar and higher curvature terms are finite.\\ From the point of view of the Finite Action Principle, all of the investigated wormhole spacetimes are included in the gravitational path integral. 
\section{Conclusions and discussion}
The Finite Action Principle is a powerful tool to study quantum gravity theories and also QFTs in general. From the point of view of finite action selection principle \cite{borissova2020blackhole,Chojnacki:2021ves} they are equally good theories, resolving the black holes singularities, assuming that ghost issue is resolved in the latter case. Yet none of the regular B-H solutions have been found in the context of H-L gravity \cite{AW} and the ones found for $f(R)$ suffer from mass inflation issue \cite{Carballo-Rubio:2018pmi,Bonanno:2020fgp,Bertipagani:2020awe}. Hence it is a strong suggestion that the wormholes may appear in the UV regime of H-L gravity and can serve as a ``cure'' for singularities \cite{Lobo:2016zle,Olmo:2016hey,Bambi:2015zch}.\\
In the case of wormholes, both traversable and non-traversable wormholes are on equal footing in the case of the Finite Action Principle, yet maybe the finite amplitude principle could distinguish between those \cite{Jonas:2021xkx}, since the amplitude to cross non-traversable wormhole should be different than the traversable one. However, this principle suggests that there is a trade-off between the resolution of black-hole singularities and the appearance of wormhole spacetimes due to higher curvature invariants. The wormhole solutions will remain in both the LI and H-L path integrals. The higher-order curvature scalars, generically present in the quantum gravity, stabilise the wormhole solutions without the need for an exotic matter.\\
\section*{Acknowledgements} We thank J. N. Borrisova, A. Eichhorn and A. Wang for inspiring discussions and careful reading of the manuscript. J.H.K. was supported by the Polish National Science Centre grant 2020/39/B/ST2/01279. J.H.K. would like to thank the CP3-Origins for the extended hospitality during this work. J.H.K acknowledges the NAWA Iwanowska scholarship PPN/IWA/2019/1/00048. 

\addcontentsline{toc}{section}{The Bibliography}
\bibliography{Ree}{}
\bibliographystyle{apsrev4-1}
\end{document}